\documentclass[aps,prr,twocolumn,amsfonts, amssymb, amsmath,showpacs,superscriptaddress,showkeys, nofootinbib,longbibliography]{revtex4-1}
\AtBeginDocument{%
\newwrite\bibnotes
\def\bibnotesext{Notes.bib}
\immediate\openout\bibnotes=\jobname\bibnotesext
\immediate\write\bibnotes{@CONTROL{REVTEX41Control}}
\immediate\write\bibnotes{@CONTROL{%
apsrev41Control,author="08",editor="1",pages="1",title="0",year="1"}}
 \if@filesw
 \immediate\write\@auxout{\string\citation{apsrev41Control}}%
\fi
}%
\usepackage[T1]{fontenc}
\usepackage[english]{babel}
\usepackage{siunitx}
\usepackage[utf8]{inputenc}
\usepackage{url}
\usepackage{dsfont}
\usepackage{bbm}
\usepackage[usenames,dvipsnames]{xcolor}
\usepackage{nicefrac}
\usepackage{setspace}
\usepackage[short]{ optidef }
\usepackage{float}
\usepackage{comment}
\usepackage[utf8]{inputenc}

\usepackage[colorlinks=true, citecolor=blue,urlcolor=blue,linkcolor=blue,filecolor=black]{hyperref}

\usepackage[colorinlistoftodos, color=green!40, prependcaption]{todonotes}
\usepackage{ulem}
\usepackage{amsthm}
\usepackage{mathtools}
\usepackage{physics}
\usepackage{xcolor}
\usepackage{graphicx}
\usepackage{adjustbox}
\usepackage{placeins}

\usepackage{lipsum}
\usepackage{csquotes}

\usepackage{multirow}   
\usepackage{pgfplots}   

\newcommand{\beq}{\begin{equation}}
\newcommand{\eeq}{\end{equation}}
\renewcommand{\emph}{\textit}

\usepackage{color,soul}

\begin{document}

\title{Quantum Entanglement Through the Lens of Paraconsistent Logic}

\author{Pouria Abbasalinejad }
\email{p.abbasalinejad@modares.ac.ir}
\affiliation{Department of Philosophy, Tarbiat Modares University, Tehran, Iran}
\author{Hamid Tebyanian}
\email{hamid.tebyanian@york.ac.uk}
\affiliation{Department of Mathematics, University of York, Heslington, York, YO10 5DD, United Kingdom}
\begin{abstract}
This paper presents an alternative approach to quantum entanglement, one that effectively resolves the logical inconsistencies without leading to logical contradictions. By addressing some of the inconsistencies within quantum mechanics, such as state superposition and non-locality, that challenge classical causal explanations, our method is constructed on the principles of paraconsistent logic. Our aim is to develop a para-consistent framework that supports the features of quantum mechanics while remaining faithful to its fundamental principles. In this pursuit, we scrutinize the philosophical and mathematical foundations of quantum mechanics in relation to classical logic systems. This method is designed to untangle theoretical puzzle spaces and promote coherence in the discussion of quantum theory. Ultimately, our approach offers a potential solution for interpreting quantum mechanics in a more coherent manner.
\end{abstract}

\maketitle

\section{Introduction}

Quantum mechanics, with its departure from classical principles, remains one of the most intellectually stimulating and controversial domains in physics. At the heart of this debate lies the phenomenon of quantum entanglement, a counterintuitive aspect wherein particles become interlinked, such that the state of one particle instantaneously influences the state of another, regardless of the distance between them. Originating from the seminal EPR paper by Einstein, Podolsky, and Rosen, which challenged the completeness of quantum mechanics \cite{S1}, entanglement was further scrutinized through Bell's theorem \cite{S2}, empirically supported by experiments conducted by Kocher \cite{S3}, Freedman and Clauser \cite{S4}, and later by Aspect et al. \cite{S5}. Interest in quantum entanglement has grown due to its potential implications for quantum information processing and computing. These systems could offer computational and cryptographic capabilities beyond the scope of classical systems \cite{S6}. Moreover, certain studies suggest the existence of correlations exceeding those anticipated by quantum mechanics, proposing a revision of its fundamental predictions \cite{S7}.

Despite its foundational role in quantum theory, entanglement poses a stark contrast to classical intuition, particularly through its challenge to the principle of locality. This paradox is manifested in the phenomenon where entangled particles exhibit correlations that, despite forbidding faster-than-light information transfer, enable instantaneous mutual information retrieval \cite{S8}. Such observations represent a profound deviation from classical logic, where measurements on one system (system A) can predict outcomes on another (system B) with uncanny precision despite no physical interaction. However, consistency is traditionally a highly important issue in physical theories. This is because, from a classical logical perspective, inconsistency leads to a theory's logical collapse. Moreover, due to precise predictions and the presence of mathematics in theories' structures, it is normal for us to be sensitive to the existence of inconsistencies in theories.

The enigma of entanglement has invited various philosophical and logical interpretations. Some approaches, drawing on metaphysical considerations, view the universe as fundamentally relational, emphasizing the ontological primacy of interconnections over isolated entities \cite{S9} \cite{S10}. Others propose frameworks based on information theory or discrete mathematical structures, suggesting entanglement reflects inherent informational principles or necessitates a hybrid logical system embracing both commutative and noncommutative elements \cite{S11} \cite{S12}.

In contrast, we posit that paraconsistent logic offers a robust framework for addressing the paradoxes inherent in quantum entanglement. Paraconsistent logic, by tolerating inconsistencies without succumbing to logical collapse, presents a viable means of reconciling quantum mechanics' inherent inconsistencies. Pioneered by logicians like Newton da Costa and Graham Priest, this approach has been applied to various quantum phenomena, offering novel interpretations of the Bohr atomic model, the complementarity principle, and quantum superposition within a paraconsistent framework \cite{S13} \cite{S14} \cite{S15} \cite{S16}. Based on this, one can consider the concept of paraconsistent from a logical perspective regarding quantum entanglement. If we accept the existence of inconsistency in quantum entanglement, its underlying logic should be paraconsistent. Paraconsistent logic can provide a logical framework for reasoning about inconsistency in quantum entanglement. Quantum entanglement by managing the inconsistencies. Considering such an approach, one can defend the inconsistencies in quantum entanglement with a new logical framework. 

This paper aims to clarify the inconsistency of quantum entanglement through the lens of paraconsistent logic. We will explore the nature of entanglement, provide a review of paraconsistent logic, and apply this logic to demystify entanglement's paradoxical essence. Thus, our analysis is structured into three core sections: an in-depth exploration of quantum entanglement, a critical examination of paraconsistent logic, and the application of paraconsistent principles in resolving the enigmas of entanglement.

\section{Quantum Entanglement}
The discussion of quantum entanglement intersects fundamentally with the measurement problem in quantum mechanics, posing the question of whether physical properties exist in a definite state prior to measurement. Consider the scenario of measuring a particle's position: did the particle possess a specific location before its measurement, or is it the act of measurement that defines its position? This dilemma splits opinions: some argue the particle's position is not predetermined, implying the particle's state is a result of measurement interactions; others suggest the particle's prior existence in a specific location, challenging quantum mechanics as incomplete for its reliance on statistical and probabilistic descriptions \cite{S17}. This latter viewpoint introduces the concept of hidden variables to account for information quantum mechanics seemingly omits.

The Einstein-Podolsky-Rosen (EPR) paradox, formulated to criticize the completeness of quantum mechanics, illustrates this debate through a thought experiment involving entangled electron and positron pairs, symbolized by a quantum state equation \(\frac{1}{\sqrt{2}}(\left|-\right\rangle\left|+\right\rangle + \left|-\right\rangle\left|+\right\rangle)\). If one measures the electron's spin and finds it to be \(\left|-\right\rangle\), it instantaneously infers the positron's spin as \(\left|+\right\rangle\), challenging the notion of faster-than-light influence as impossible by classical standards \cite{S18}.

Einstein, Podolsky, and Rosen did not refute quantum mechanics but argued for its incompleteness, suggesting that the wave function alone does not provide a full account of physical reality. Efforts to incorporate hidden variables into quantum theory emerged, but in 1964, John Bell's theorem exposed a fundamental incompatibility between hidden variables and the intrinsic non-locality of quantum mechanics. Bell's theorem, extending the EPR paradox, highlighted entanglement as a distinctive quantum feature, absent in classical mechanics, where the measurement of one particle immediately reveals the state of its entangled counterpart, irrespective of distance.

Empirical validations of Bell's theorem and the exploration of entanglement continue, yet interpretations vary, especially philosophically and logically, due to quantum mechanics' interpretive diversity. This paper endorses the Copenhagen interpretation, one of several, which dismisses the pre-measurement existence of particle properties as inconsequential. It posits that measurement by an external observer or device triggers wave function collapse, with the wave function serving as a mathematical construct for calculating event probabilities and not as a direct physical description \cite{S19}. The Copenhagen interpretation, embracing quantum indeterminacy, asserts nature's description as inherently probabilistic, underpinned by the Heisenberg uncertainty principle, the Bohr complementarity principle showcasing wave-particle duality, and the correspondence principle, aligning quantum mechanics with classical physics on a macroscopic scale. The Copenhagen interpretation argues against counterfactual definiteness and the necessity for hidden variables, maintaining quantum mechanics' status as an epistemic theory rather than offering a definitive description of reality \cite{S20}.

To reconcile the contrast between quantum entanglement and classical understanding, we must thoroughly investigate the inherent inconsistencies. In the following sections, we will delve into this matter, aiming to enhance the discussion on quantum entanglement by incorporating mathematical formulations

\section{Inconsistency in Quantum Entanglement}

The principle of quantum superposition, which allows a quantum system to exist in multiple states simultaneously until measured, presents a fundamental departure from classical physics and illustrates the first major inconsistency. In classical physics, the superposition of waves or fields can be represented mathematically as:

\begin{equation}
    \mu = \alpha + \beta
\end{equation}

where \(\mu\) is the resulting wave or field from the superposition of \(\alpha\) and \(\beta\). In contrast, quantum mechanics describes the state of an entangled system in superposition as:

\begin{equation}
     |\phi\rangle = |a\rangle |b\rangle + |a'\rangle |b'\rangle
\end{equation}

This equation denotes that, prior to measurement, the system simultaneously resides in both states \(|a\rangle |b\rangle\) and \(|a'\rangle |b'\rangle\). The act of measurement collapses the system into one of these states, illustrating the probabilistic nature of quantum mechanics.

The mathematical essence of superposition in quantum mechanics can be further understood through the linearity of the Schrödinger equation. If \(x\) and \(y\) are solutions to the Schrödinger equation, then any linear combination \(ax + by\) will also be a solution, as per the principle of superposition:

\begin{equation}
     \Psi = ax + by
\end{equation}

This principle implies that an object in a quantum state is not confined to a single state but exists in all possible states simultaneously until observed. This is starkly contrasted with classical physics, where the future state of a system can be determined by its initial conditions.

The concept of non-locality, which allows entangled particles to influence each other’s states instantaneously regardless of distance, introduces another layer of inconsistency that challenges the classical causality principle of cause and effect being local and sequential. Quantum entanglement suggests that a change in the state of one particle instantaneously affects its entangled partner, regardless of the distance between them, which can be described using Bell's inequality:

\begin{equation}
    S = |E(a, b) - E(a, b')| + |E(a', b) + E(a', b')| \leq 2 
\end{equation}

However, quantum experiments have consistently produced results where \(S > 2\), indicating a violation of Bell's inequality and thus confirming the non-local nature of quantum entanglement.

Quantum mechanics, with its inherent probabilistic and non-local characteristics, necessitates a departure from deterministic classical models, demanding a new logical framework to reconcile these differences.

Following the introduction of Bell's inequality, it becomes evident that quantum mechanics permits scenarios that defy classical expectations. The essence of Bell's theorem, through its foundational inequality, challenges the very fabric of locality and causality — concepts that are cornerstones of classical physics. The empirical violations of this inequality not only corroborate the non-local nature of quantum entanglement but also underscore the inadequacy of local realism to fully capture the phenomena inherent in quantum mechanics. 
To better understand this inconsistency, let us talk about CHSH (Clauser-Horne-Shimony-Holt) inequality. The CHSH game represents a scenario used to test the limits of quantum entanglement and non-locality, contrasting sharply with classical expectations. The game involves two players, Alice and Bob, who are separated and cannot communicate. They each receive a binary input (0 or 1) and must produce a binary output (0 or 1) based on the inputs they receive. The goal is for the sum of their outputs, modulo 2, to equal the product of their inputs:
\[ (output\_A \oplus output\_B) = (input\_A \cdot input\_B) \]
Alice and Bob share an entangled pair of qubits in a quantum setting. The quantum strategy surpasses classical limits using specific measurement bases depending on their inputs. The optimal quantum strategy involves measurements in the following bases:
\begin{itemize}
    \item If \(input\_A = 0\), Alice measures in the \(Z\) basis.
    \item If \(input\_A = 1\), Alice measures in the \(X\) basis.
    \item If \(input\_B = 0\), Bob measures in the \( \frac{Z + X}{\sqrt{2}} \) basis.
    \item If \(input\_B = 1\), Bob measures in the \( \frac{Z - X}{\sqrt{2}} \) basis.
\end{itemize}
Using these settings and sharing an entangled state such as the Bell state \( \frac{1}{\sqrt{2}}(|00⟩ + |11⟩) \), they can achieve a success probability exceeding classical limits, specifically up to approximately $85.4 \%$ (or \(\cos^2(\pi/8)\)), compared to $75 \%$ for the best classical strategy.

The CHSH inequality is a quantitative measure of the degree to which quantum mechanical experiments disagree with local hidden variable theories. It is expressed as \(|E(a, b) - E(a, b') + E(a', b) + E(a', b')| \leq 2 \).
Here, \(E(a, b)\) represents the expectation value of the product of outcomes on settings \(a\) and \(b\). Using the measurement above settings and the entangled Bell state, the quantum version of this expectation can reach up to:
\( 2\sqrt{2} \),
known as Tsirelson's bound, a value that surpasses the classical limit of 2. This starkly demonstrates a fundamental departure from classical expectations, thereby expressing non-locality \cite{S21} \cite{S22} \cite{S23}

The violation of the CHSH inequality highlights a type of logical conflict that traditional binary logic cannot resolve, yet it naturally manifests in quantum contexts. In these scenarios, paraconsistent logic, which accommodates inconsistencies, proves helpful. It provides a coherent approach to managing situations that would otherwise lead to unacceptable inconsistencies in classical logic frameworks, such as those observed in quantum entanglement scenarios exemplified by the CHSH game.

When dealing with entangled systems, we encounter two types of inconsistency: the inconsistency arising from the "Non-locality problem" and the "superposition problem". This revelation prompts a deeper examination of the principles underpinning quantum theory, revealing a realm where particles maintain a connection that transcends spatial separation and conventional causality. It's this aspect of quantum mechanics that necessitates a reevaluation of our classical intuitions, inviting us into a nuanced understanding of reality where distance becomes irrelevant to the instantaneous relational dynamics between entangled particles.

\section{Paraconsistent logic}

Paraconsistent logic challenges the traditional Principle of Explosion, or Ex Contradictione Quodlibet (ECQ), which holds that from a contradiction, any statement can be derived, denoted as \( \alpha, \neg \alpha \vdash \beta \), (\(\neg\) is a negation operator and \(\vdash\) is a logical consequence). Paraconsistent logic sets aside this principle (\( \alpha, \neg \alpha \nvdash \beta \)), allowing for the coexistence of inconsistencies without leading to the trivialization of the system \cite{S24}. This logic is instrumental in situations where inconsistent information arises, yet reasoning must continue sensibly. Paraconsistency serves as a foundational condition for logical systems to accommodate inconsistencies without succumbing to triviality \cite{S25}. The primary motivation behind paraconsistent logic is the recognition that in certain scenarios, we encounter information or theories that are inconsistent, yet it remains crucial to continue reasoning in a sensible manner. This perspective challenges Popper’s assertion about the detrimental impact of accepting inconsistency on scientific theories, suggesting instead that paraconsistent logic can effectively manage inconsistencies within a system \cite{S26}. The existence of consistency in a theory's foundational logic implies that no theory can be proven to entail every sentence. In that case, we would be faced with a theory that entails everything, leading to its triviality. Therefore, consistency is important in this regard, but the logical basis of this triviality lies in the principle of explosion, and this issue is not raised in paraconsistent systems.

Graham Priest believes that throughout the history of Western philosophy, numerous individuals have emphasized paraconsistent views \cite{S27}. He considers the oldest of these individuals to be pre-Socratic philosophers such as Heraclitus. Some medieval Neoplatonic philosophers like Nicholas of Cusa have had inconsistent theories. Priest also refers to Hegel, among later philosophers. However, Jan Łukasiewicz and Nicolai Vasiliev independently made efforts in 1910 and 1911, respectively, and Stanisław Jaśkowski in 1948. This approach was further developed by Florencio González Asenjo in 1954 and Newton da Costa in Brazil in 1963. Asenjo proposed the first multi-valued inconsistent logic, and da Costa presented the principle-based systems for a specific family of paraconsistent logics and formulated the first paraconsistent logic.

Within the framework of paraconsistent logic, various logical systems have been proposed that share the common feature of abandoning the principle of explosion, which states that any statement can be derived from a contradiction. In this article, we will focus on "Logics of Formal Inconsistency" (LFIs) or \(C_{n}\) logics introduced by da Costa (we need the first order of this logical system, \(C_{1}\)). \(C_{n}\) logics are a family of paraconsistent logics that incorporate certain parts of classical logic while rejecting the principle of explosion where inconsistencies occur. In this approach, we can explicitly separate inconsistency from triviality and study inconsistent theories without assuming they are necessarily trivial. This demonstrates that the existence of a substantive Inconsistency is distinct from the trivial nature of inconsistent theories. 
da Costa proposes several constraints to elucidate a paraconsistent logic. One of them is that paraconsistent logic should not include the principle of explosion, represented by the formula \(\neg(\alpha \land \neg\alpha)\), as a logical truth. Graham Priest argues that the idea behind this is that if someone has a theory containing propositions \(\alpha\) and \(\neg\alpha\), they do not want a logical truth that connects these two propositions \cite{S27}. Looking at da Costa's later works, one can agree with Priest's opinion. Another constraint proposed by da Costa is that logic must include classical logic or at least intuitionistic logic. That is, he says that classical logic should limit paraconsistent logic to a certain extent. Priest considers this condition overly stringent. However, we believe that paraconsistent logic leading to classical logic under specific conditions strongly resembles the classical limit in quantum mechanics, which would be suitable for advancing our work. Now, we discuss the syntactic and semantic structure of logic \(C_{1}\) \cite{S28}.

\subsubsection*{Syntactic Structure}

\begin{description}
  \item [Vocabulary of \(C_{1}\)]. \\
    Propositional variables: An uncountable (infinite) set of variables. \\   
    Logical operators: \(\neg\), \(\lor\), \(\land\), \(\rightarrow\) and \(\leftrightarrow\). \\
    Punctuation symbols: Open and close parentheses.
    
  \item[Construction Rules of \(C_{1}\)]. \\ 
    Every propositional variable is a formula. \\
    If \(\alpha\) and \(\beta\) are formulas, then \(\neg \alpha\), \(\neg \beta\), \(\alpha \lor \beta\), \(\alpha \land \beta\), \(\alpha \rightarrow \beta\) and \(\alpha \leftrightarrow \beta\) are also formulas. \\
    For every formula in \(C_{1}\), the principle of non-contradiction must generally hold, and no proposition can be inferred from two contradictory propositions.

  \item[Definitions of \(C_{1}\)]. \\ 
    \(\alpha \leftrightarrow \beta = (\alpha \rightarrow \beta) \lor (\beta \rightarrow \alpha)\) \\
    \(\alpha_0 = \neg (\alpha \land \neg \alpha)\) \\
    \(\neg^*\alpha = \neg \alpha \land \alpha_0\)

  \item[Postulates of \(C_{1}\)]. \\ 
    \(\alpha \rightarrow (\beta \rightarrow \alpha)\) \\
    \((\alpha \rightarrow \beta) \rightarrow ((\alpha \rightarrow (\beta \rightarrow \gamma)) \rightarrow (\alpha \rightarrow \gamma))\) \\
    \(\alpha, \alpha \rightarrow \beta \quad / \quad \therefore \beta\) (Modus Ponens (MP)) \\
    \(\alpha \land \beta \rightarrow \alpha\) \\
    \(\alpha \land \beta \rightarrow \beta\) \\
    \(\alpha \rightarrow (\beta \rightarrow \alpha \land \beta)\) \\
    \(\alpha \rightarrow (\alpha \lor \beta)\) \\
    \(\beta \rightarrow (\alpha \lor \beta)\) \\
    \((\alpha \rightarrow \gamma) \rightarrow ((\beta \rightarrow \gamma) \rightarrow (\alpha \lor \beta \rightarrow \gamma))\) \\
    \(\beta_0 \rightarrow ((\alpha \rightarrow \beta) \rightarrow (\alpha \rightarrow \neg \beta) \rightarrow \neg \alpha)\) \\
    \(\alpha_0 \land \beta_0 \rightarrow (\alpha \land \beta)_0 \land (\alpha \lor \beta)_0 \land (\alpha \rightarrow \beta)_0\) \\
    \(\alpha \lor \neg \alpha\) \\
    \(\neg \neg \alpha \rightarrow \alpha\)
\end{description}

\subsubsection*{Semantic Structure}

\begin{description}
\item[Interpretation of Language \(C_{1}\)] : \\
    We consider two-valued semantics, with truth 1 and falsehood 0 for \(C_{1}\).\\
    The interpretation function \(V = F \rightarrow \{0,1\}\) assigns values to formulas.
    
\item [Validity in an Interpretation].\\
    \(\models\alpha\): The interpretation function \(V\) assigns the value \(1\) to the formula \(\alpha\), making it valid.\\
    \(\not\models\alpha\): The interpretation function \(V\) assigns the value \(0\) to the formula \(\alpha\), making it invalid.
    
    \item[Semantic Rules]. \\
    \(V(\alpha)=0 \Longrightarrow V(\neg\alpha)=1\).\\
    \(V(\neg\neg\alpha)=1 \Longrightarrow V(\alpha)=1\).\\
    \(V(\beta_{0}) = V(\alpha\rightarrow\beta) = V(\alpha\rightarrow\neg\beta)=1 \Longrightarrow V(\alpha)=0\).\\
    \(V(\alpha\rightarrow\beta)=1 \Longleftrightarrow V(\alpha)=0 \: or \: V(\beta)=1\).\\
    \(V(\alpha\land\beta)=1 \Longleftrightarrow V(\alpha)=V(\beta)=1\).\\
    \(V(\alpha\lor\beta)=1 \Longleftrightarrow V(\alpha)=1 \: or \: V(\beta)=1\).\\
    \(V(\alpha_{0})=V(\beta0)=1 \Longrightarrow V((\alpha\lor\beta)_{0})=V((\alpha\land\beta)_{0})=V((\alpha\rightarrow\beta)_{0})=1\)
\end{description}

After reviewing this logical system's syntactic and semantic structure, we can move on to its metatheory. Regarding the relationship between the syntactic and semantic structure of \(C_{1}\), both the soundness metatheorem (\(\Gamma\vdash \Longrightarrow \Gamma\models\alpha\)) and the completeness metatheorem (\(\Gamma\models \Longrightarrow \Gamma\vdash\alpha\)) hold true. Additionally, propositional logic \(C_{1}\) is a decidable logic.

Here, attention must be paid to the Law of Non-Contradiction (LNC), i.e., \(\vdash \neg(\alpha\land\neg\alpha)\) (\(\models \neg(\alpha\land\neg\alpha)\)). The syntactic and semantic structure of C1 is such that despite abandoning ECQ, it still adheres to LNC and does not accept the existence of contradictions in the system. That is, the existence of inconsistency in the system does not mean accepting contradictory propositions. So, in this logical system, if we have \( T \vdash \alpha \) and \( T \vdash \neg\ \alpha\ \) (\(T\) is a theory), it does not necessarily mean that we can derive \( T \vdash \neg\ \alpha\ \wedge \alpha \) \cite{S29}. Accordingly, if \(\alpha \) is a proposition in the language of theory \(T\) (for example, quantum theory),  in general, \(\alpha\ ,\neg\ \alpha \vdash \alpha\ \wedge \neg \alpha \) does not hold. Also, considering eliminating the explosion principle, we will have \(\nvdash (\alpha \wedge \neg \alpha) \rightarrow \eta\).

This means we cannot have the adjunction of two inconsistent propositions simply because they exist. Even if there is an adjunction of two incompatible propositions, we cannot conclude an arbitrary proposition, which means that we have prevented the principle of explosion. So, In da Costa's paraconsistent logic \(C_{n}\), the presence of a proposition $\alpha$ and its negation $\neg \alpha$ does not lead to the automatic derivation of any conclusion \(\beta\) , formally $\{\alpha, \neg \alpha\} \nvdash \beta$. This principle is crucial for maintaining the coherency of logical systems in the face of inconsistencies. Paraconsistent logic, therefore, provides a foundation for reasoning within systems containing inconsistent information without succumbing to the logical extremism of the Principle of Explosion.

\section{Paraconsistent analysis of quantum entanglement}

As mentioned earlier, one of the inconsistencies observed in quantum entanglement is the issue of quantum superposition. The application of paraconsistent logic extends to the interpretation of quantum entanglement, particularly in addressing the paradox of quantum superposition. Da Costa and De Ronde's approach to quantum superposition within a paraconsistent framework suggests a reevaluation of quantum mechanics \cite{S16}. Assuming a system including an electron and a positron in the state \(\frac{1}{\sqrt{2}}( \left| - \right\rangle\left| + \right\rangle + \left| - \right\rangle\left| + \right\rangle )\), the Copenhagen interpretation posits that before measurement, particles exist in simultaneous inconsistent states. 

However, the mathematics used in quantum mechanics and formulating entanglement is based on classical logic. Therefore, considering that we aim to transition the foundational logic from classical to paraconsistent, the desired mathematics should also be consistent with paraconsistency. If we denote classical mathematics as CM and paraconsistent mathematics as PM, we call the developed quantum mechanics in CM and PM as QM and PQM, respectively. Here, we intend to refrain from rewriting the mathematics of quantum mechanics, as this is a vast project beyond this article's scope. The sole purpose of stating this matter is to understand that if we intend to use paraconsistent logic as the logical foundation of quantum mechanics, we must also pay attention to its mathematics. Now, considering that paraconsistent logic encompasses classical logic, all statements proven in CM will also be provable in PM.
Consequently, all QM formulas will also be provable in PQM. In fact, by changing the paraconsistent logical framework and subsequently changing the mathematics, we do not lose anything from quantum mechanics. In other words,
\begin{equation}
    CL \subseteq PL \Longrightarrow CM\subseteq PM \Longrightarrow QM \subseteq PQM
\end{equation}

Now, considering \( \alpha\) as a proposition representing the state of an entangled system composed of an electron and positron, according to the principle of superposition, the entangled system within the framework of PM will be as follows:

\begin{itemize}
    \item \(\alpha_{1}\): The entangled system is in the state $\left| - \right\rangle\left| + \right\rangle$.
    \item \(\neg \alpha_{1}\): The entangled system is not in the state $\left| - \right\rangle\left| + \right\rangle$.
    \item \(\alpha_{2}\): The entangled system is in the state $\left| + \right\rangle\left| - \right\rangle$.
    \item \(\neg \alpha_{2}\): The entangled system is not in the state $\left| + \right\rangle\left| + \right\rangle$.
\end{itemize}

Therefore, the concept of superposition can be written as a conjunction of states \(\neg \alpha_{1} \wedge \alpha_{1} \wedge \neg \alpha_{2} \wedge \alpha_{2} \) \cite{S16}. However, considering \(\nvdash (\alpha \wedge \neg \alpha) \rightarrow \eta\) in paraconsistent logic, it can be said that despite the existence of inconsistency, the system will still be logically acceptable because this theorem prevents the collapse of the system and the transformation of quantum mechanics into a trivial theory.

So far, the inconsistent aspects of superposition have been addressed in entanglement. Now, we need to focus on the inconsistency between locality and non-locality. To address the apparent inconsistency between instantaneity in quantum entanglement and the principle of locality, we consider two propositions:

\begin{itemize}
    \item \(\beta_{1}\) (Proposition 1): According to the principle of locality, any physical system (e.g., an electron) is influenced by its immediate surroundings (e.g., a positron) at the speed of light
    \item \(\beta_{2}\) (Proposition 2): In an entangled system comprising an electron and a positron, information transfer between the particles occurs instantaneously, i.e., at a speed surpassing that of light.
\end{itemize}
This scenario presents a complementarity in quantum entanglement, indicating that if a theory \(T\) incorporates both descriptions \(\beta_{1}\) and \(\beta_{2}\), which belong to the same language but neither can singly account for the phenomena, their combined description leads to a logical inconsistency \cite{S30}. Specifically, if \(T\) contains both true formulas \(\beta_{1}\) and \(\beta_{2}\), and their conjunction from a classical logic standpoint results in inconsistency. Using da Costa's method, we denote the quantum mechanics theory as \(T\), with \(\beta_{1}\) and \(\beta_{21}\) as theorems within \(T\), thus: \(T \vdash \beta_{1}\) and \(T \vdash \beta_{2}\).

Now, we want to interpret this inconsistency using da Costa's method \cite{S15}. For this purpose, we designate the quantum mechanics theory as T and consider its propositions \(\beta_{1}\) and \(\beta_{2}\) as \(T \vdash \beta_{1}\) and \(T \vdash \beta_{2}\), respectively. That is:

\[T, \beta_{1}\vdash \gamma \space\ \space\ and \space\ \space\ T, \beta_{2}\vdash \neg \gamma \].

In general, \(\gamma\) indicates that locality is established, while \(\neg \gamma\) signifies that locality is not established (we have non-locality). Adopting paraconsistent logic in lieu of classical logic for \(T\), despite the presence of inconsistency, allows \(T\) to retain logical coherence. In this system, from \(T, \beta_{1}\vdash \gamma\) and \(T, \beta_{2}\vdash \neg \gamma\), it does not necessarily follow that \(T \vdash \gamma \wedge \neg \gamma\). Paraconsistent logic includes the theorem \(\nvdash p \gamma \wedge \neg \gamma \rightarrow \eta\), suggesting that even with \(\gamma \wedge \neg \gamma\), \(T\) avoids collapse into triviality, maintaining its logical coherence. Here, we will face two complementary propositions leading us to the complementarity of locality and non-locality. Similar to the complementarity of particle and wave, in entangled systems, we encounter the complementarity of locality and nonlocality in this context.

Paraconsistent logic enables a unique interpretation in entanglement scenarios, where measurement outcomes on one particle instantaneously affect another regardless of distance. It allows outcomes to be considered isolated and connected simultaneously, avoiding the classical logical impasse that would typically result in a breakdown of logical consistency. This logic also supports the notion of superposition, where particles exist in multiple states at once. Paraconsistent logic allows these states to coexist without forcing a collapse into a single state upon measurement, thus maintaining the integrity of quantum descriptions and aligning well with behaviours observed in particles like electrons and photons, which exhibit characteristics of both particles and waves.

Classical logic, which strictly adheres to the Principle of explosion, can only accommodate these by falling into logical absurdity. In contrast, paraconsistent logic, which tolerates inconsistencies and rejects the Principle of explosion, offers a fresh perspective on these quantum phenomena. The adaptability of paraconsistent logic to quantum mechanics is incredibly beneficial. It allows theorists to explore the implications of quantum phenomena freely without being restricted by classical logical paradigms that would necessitate discarding valid observational data due to perceived inconsistencies. Paraconsistent logic acknowledges the coexistence of inconsistencies truths, a fundamental characteristic of quantum mechanics, and is crucial for interpreting entangled states. These states exhibit dependent behaviours suggesting a pre-existing connection despite properties like spin, position, or momentum not being definitively assigned until measurement.
The application of paraconsistent logic to quantum mechanics extends beyond theoretical novelty. It offers practical implications for understanding information sharing and preservation within a quantum system. For instance, in quantum computing, the manipulation of entangled qubits relies on their non-classical correlation, which classical logic cannot adequately describe without encountering paradoxes.

\section{paraconsistent friendly CHSH game}

Aforementioned, entanglement between two particles, \(A\) and \(B\), can be mathematically represented by the state vector 
\(
|\psi\rangle = \alpha|00\rangle + \beta|11\rangle
\)
where \(|00\rangle\) and \(|11\rangle\) represent the states where both particles are simultaneously in states 0 and 1 respectively, and \(\alpha\) and \(\beta\) are complex coefficients that satisfy the normalization condition \(|\alpha|^2 + |\beta|^2 = 1\). This state vector describes a maximally entangled system where measurements on one particle instantaneously affect the state of the other, irrespective of the distance between them.

The propositions \(p: "A \text{ is in state } |0\rangle"\) and \(q: "A \text{ is in state } |1\rangle"\) traditionally cannot both be true in classical logic as they represent mutually exclusive states. However, within a paraconsistent framework, these propositions are assigned truth values that reflect the quantum mechanical reality of superposition. The paraconsistent truth table could be structured as follows:
\[
\begin{array}{c|c}
\text{Proposition} & \text{Truth Value} \\
\hline
p & B \\
q & B \\
\end{array}
\]
Here, \(B\) symbolizes the unique status of quantum superpositions where a quantum state can indeed embody inconsistent properties without leading to logical absurdity, embodying paraconsistency where inconsistencies are both recognized and sustained.

Logical connectives need redefinition to accommodate inconsistencies inherent in quantum mechanics:
\begin{itemize}
    \item The conjunction \(p \land q\) in paraconsistent terms does not reduce to false despite \(p\) and \(q\) being classically contradictory. Instead, it is represented as \(B\), acknowledging the dual truth of \(p\) and \(q\) reflective of the entangled or superposed state.
    \item The disjunction \(p \lor q\) is treated similarly, where the presence of either state being true in any context is sufficient, marked as \(T\) or \(B\).
\end{itemize}

The observable operators \( \hat{A} \) and \( \hat{B} \), associated with measurements on particles \(A\) and \(B\), determine the outcomes through:
\begin{equation}
    E(\hat{A}, \hat{B}) = \langle \psi | \hat{A} \otimes \hat{B} | \psi \rangle
\end{equation}

This expectation value, calculated through standard quantum mechanics, can exceed the limits set by classical physics, such as those outlined by Bell’s inequality. Within the paraconsistent framework, these exceeding values (\(E(\hat{A}, \hat{B})\)) are not anomalies but are indicative of the non-classical, interconnected reality of entangled states, supporting the idea that both particles can exist in inconsistent states simultaneously.

In the CHSH game, Alice and Bob use specific rotational angles for their measurements. These angles can be seen as degrees of “truth” and “falsity” in propositions that can be both true and false at the same time. The CHSH value, derived from these angles, serves as a measure of the “degree of inconsistency” in this logic system. A higher CHSH value signifies a greater degree of inconsistency, which mirrors the inherently inconsistent nature of quantum superposition and entanglement.
In the game, Alice and Bob receive random bits and must return new bits so that a certain formula holds true. Their best strategy is to measure in the basis they’re challenged with and respond with the outcome. When Bob measures, the desired outcome vector is at an angle of $22.5^{\circ}$ from his state.
Let’s relate this to paraconsistent logic. We can use propositions (p) and (q) to denote the states of Alice and Bob’s qubits. In quantum superposition, both qubits are in a mixed state, meaning that both (p) and (q) are simultaneously true and false, i.e., (p = B) and (q = B), where (B) stands for both true and false.
The rotational angles in the CHSH game can be viewed as representing the degree of “truth” and “falsity” in these paraconsistent propositions. For instance, a rotation of $0^{\circ}$ might represent “completely true”, a rotation of $180^{\circ}$ might represent “completely false”, and a rotation of $22.5^{\circ}$ (as in Bob’s optimal strategy) might represent a state that is both somewhat true and somewhat false.
The CHSH value, which can be computed from the measurement outcomes and hence indirectly from the rotational angles, can be interpreted as a measure of the “degree of inconsistency” in the paraconsistent logic system. A higher CHSH value would indicate a greater degree of inconsistency.

\begin{figure*}
\centering
\includegraphics[width=\linewidth]{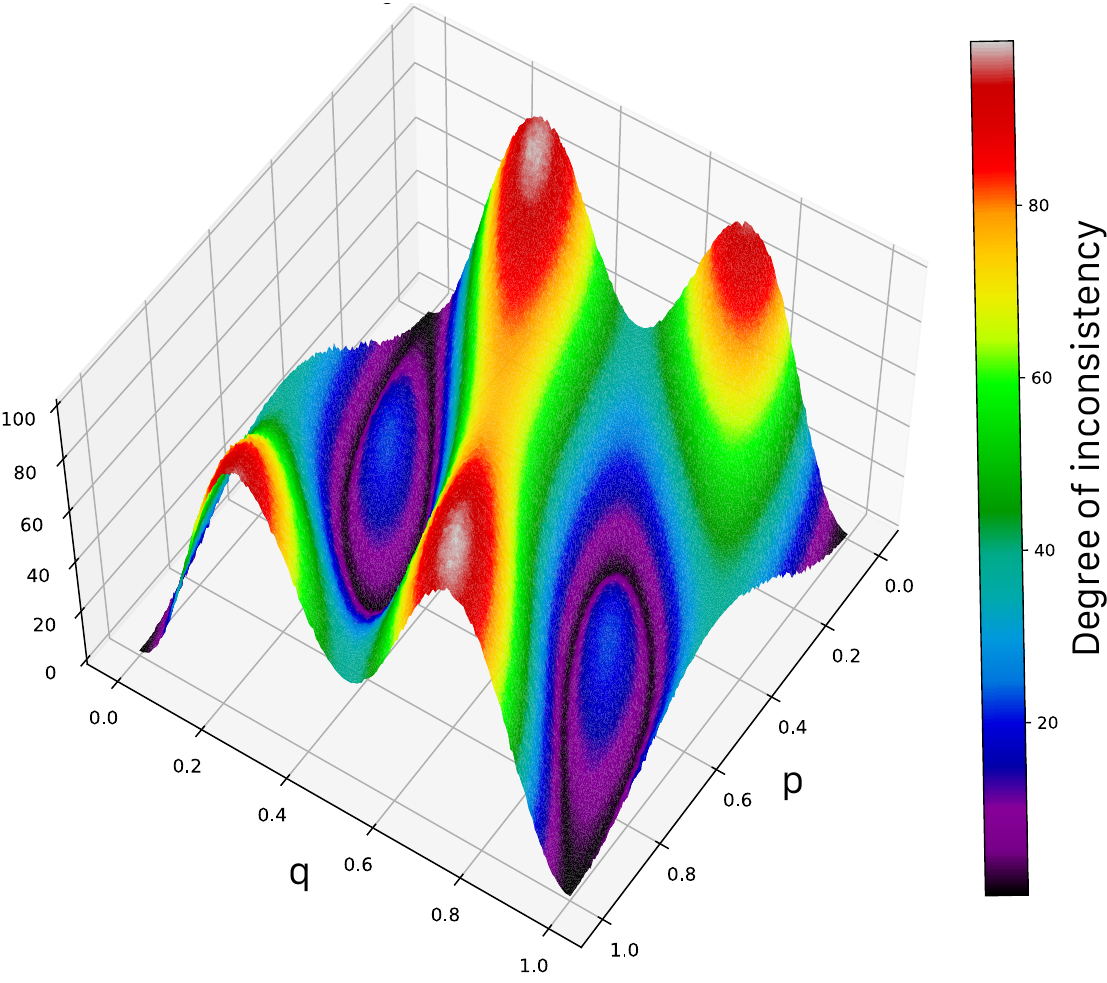}
\caption{3D representation of the degree of inconsistency in the CHSH game as a function of measurement angles for Alice and Bob. The x-axis and y-axis represent the normalized measurement angles for Alice and Bob, respectively, scaled to lie between 0 and 1. The z-axis represents the degree of inconsistency, calculated as the absolute difference between the CHSH value and the classical limit of 2, and rescaled to lie between 0 and 100. This visualization demonstrates how the degree of inconsistency varies with the measurement angles, providing a graphical interpretation of the inherently inconsistent nature of quantum superposition and entanglement.}
\label{fig:res}
\end{figure*}

Building on the previously described theoretical framework, Figure \ref{fig:res} illustrates the degree of inconsistency in the CHSH game. This figure maps the measurement angles of Alice and Bob to degrees of "truth" and "falsity" within a paraconsistent logic system using the IBM quantum computer. The measurement angles are normalized: Alice's are plotted along the x-axis and Bob's along the y-axis, each ranging from 0 to 1. The z-axis measures the degree of inconsistency, computed as the absolute difference between the CHSH value and the classical limit of 2, and rescaled to range from 0 to 100. This graphical representation allows for a clear visual interpretation of the quantum paradoxes expressed in terms of paraconsistent logic.

The surface in Figure \ref{fig:res}'s colour gradient visually signifies the degree of inconsistency. This design aids in comprehending how the inconsistency levels fluctuate with varying measurement angles, offering a visual interpretation of the paradoxical nature of quantum superposition and entanglement.

The data in Figure \ref{fig:res} not only supports our theoretical analysis but also has profound practical implications. It reveals that the degree of inconsistency can surpass the classical boundary under specific measurement configurations. This evidence underscores the immediate and tangible utility of paraconsistent logic in analyzing quantum mechanical systems, capturing the inherent inconsistencies presented by quantum entanglement and superposition.

\section{A discussion on paraconsistent approaches to quantum entanglement}

This study aimed to present a logical interpretation of quantum entanglement and clarify the inconsistencies in an entangled system. Along this path, we showed that quantum entanglement can be interpreted within a logical framework using paraconsistent logic developed by da Costa. This logical interpretation agrees with the Copenhagen interpretation of quantum mechanics. Firstly, a logical paraconsistent system is reduced to classical logic, similar to the Correspondence Principle in the Copenhagen interpretation, where classical limits are somehow respected in logic. In other words, if we consider suitable logic for quantum mechanics as paraconsistent logic and classical mechanics' logical foundation as classical logic, then, according to the Correspondence Principle, paraconsistent logic should reduce to classical logic at the classical limit. Paraconsistent logic C1 has this property and can be reduced to classical logic. On the other hand, paraconsistent logic provides the possibility of logical support for the principle of complementarity concerning locality and non-locality in the issue of entanglement. Accepting complementarity propositions for quantum mechanics is logically possible with greater precision.

\begin{figure}[htbp]
  \centering
  \tikzset{>=latex}
  \begin{tikzpicture}[scale=1]
    \def\height{-1.4}
    \def\width{2.5}
    \draw[->,thick,yshift=8pt,scale=1]
      (1.5*\width,0) -- (2.7*\width,0)
      node[above=0pt,above right,pos=0.] {Classical Limit}; 
    \draw[thick]
      (0,0) rectangle (3*\width,2*\height);
    \draw[thick]
      (0,0) rectangle (\width,\height)
      node[pos=.5,align=center] {Physics};
    \draw[thick]
      (0,\height) rectangle (\width,2*\height)
      node[pos=.5,align=center] {Logic};
    \draw[thick]
      (\width,\height) rectangle (2*\width,2*\height)
      node[pos=.5,align=center] {Paraconsistent\\Logic};
    \draw[thick]
      (2*\width,\height) rectangle (3*\width,2*\height)
      node[pos=.5,align=center] {Classical\\Logic};
    \draw[thick]
      (\width,0) rectangle (2*\width,\height)
      node[pos=.5,align=center] {Quantum\\Mechanics};
    \draw[thick]
      (2*\width,0) rectangle (3*\width,\height)
      node[pos=.5,align=center] {Classical\\Mechanics};
  \end{tikzpicture}
  \caption{The principle of correspondence in physics and logic for the classical limit.}
\end{figure}

Certainly, there may be disagreements regarding whether the claimed inconsistencies exist. However, what is clear is that observations and theories to date indicate that, at least in some cases, such as quantum entanglement, we are faced with the propositions of both \(\alpha\) and \(\neg\alpha\). However, generally, when confronted with theoretical inconsistencies, two approaches can be adopted: "consistency preservation" or "inconsistency toleration" \cite{S31}.

In the first approach, emphasis is placed on preserving the consistency of the theory, and when faced with inconsistencies, it can involve compartmentalization and information restriction. In this approach, the theory (in this case, quantum mechanics) should never involve any inconsistencies, and a theory's content and empirical success should be based on logical consistency. Essentially, the necessary condition for any theory is the absence of inconsistencies, as inconsistent theories cannot provide valid descriptions of scientific phenomena. The logical reason behind this is that, according to classical logic, from \(\alpha\land\neg\alpha\), we can derive \(\alpha\) and \(\neg\alpha\), and consequently \(\alpha\lor\beta\), ultimately leading to any arbitrary proposition \(\beta\). Therefore, inconsistency leads to any proposition's implication, necessitating the theory's triviality. Since this perspective interprets scientific theories within the framework of classical logic, any inconsistency inevitably leads to such consequences. It is observed that what renders a theory useless and trivial is not the presence of inconsistency but rather the explosion principle, which allows us to infer any proposition from inconsistency. Therefore, if we speak of the existence of inconsistency in the issue of quantum entanglement, it does not mean that our theory is insignificant. Instead, we aim to find a way to accept inconsistency in the theory without logical collapse to enhance its conformity with empirical observations. This leads us to the second approach, namely inconsistency tolerance.

In the second approach, based on a logic-driven method \cite{S32}, logical support for the existing inconsistency in the theory can be provided, facilitating the peaceful coexistence of inconsistencies in the theory's structure. That is, when faced with inconsistency, by changing the theory's underlying logic, we prevent the logical collapse of the theory. In other words, this approach aims to preserve theories logical coherence and implications. This is precisely the method we employed in examining the issue of quantum entanglement. Instead of rejecting the theory when faced with inconsistency, we modify its logical framework. However, this does not imply that quantum mechanics is contradictory. Because in the logical system we utilized, two inconsistent propositions cannot generally lead to a contradiction, i.e., \(\nvdash \alpha\land\neg\alpha\). Paraconsistent logic provides the possibility for the issue of entanglement to maintain a logical foundation despite the existence of inconsistent propositions, guiding us to provide a logical framework for the issue of locality complementarity.

Of course, regarding the issue of locality complementarity, it is necessary to pay attention to an important point. The principle of locality can be viewed in two ways. Either we consider it a consequence of Einstein's theory of relativity, which has no connection to quantum mechanics, or we consider it a principle belonging to quantum mechanics (or physics as a whole). If we consider the second case, we are faced with a physical principle called locality, which belongs to the language of quantum mechanics. However, empirical observations and theoretical results indicate that particles exhibit local and non-local behaviors. Similar to the complementarity principle regarding wave-particle duality, we can consider "locality complementarity" for this issue. That is, logically, we are faced with two propositions: \(\gamma\) (locality governs a physical system) and \(\neg\gamma\) (locality does not govern a physical system).
This inconsistency might seem more important and challenging due to its entrapment with causality. We believe that the roots of such challenges lie in our intuitions and classical logic. Therefore, if we deal with the issue based on classical logic, the existence of inconsistency would lead to the collapse of the theory, and logically, we would have to discard it and turn to another theory. However, at least so far, the physics community has yet to be willing to do so. This is because empirical observations show that a physical system violates the principle of locality under certain conditions. We believe nothing transcends experience and experimentation, even if it leads to abandoning our deepest classical intuitions.
For this reason, we suggest considering the term "strong inconsistency" for the inconsistency between locality and non-locality for the issue of entanglement. However, using the perspective of paraconsistency, if the theory's underlying logic is paraconsistent, this inconsistency, even if it arises in the context of locality and causality, does not lead to the collapse of the theory. However, this does not mean we should discard the concept of causality because of this issue. Here, we are faced with a duality regarding the principle of locality, and just as the wave-particle duality does not imply giving up one for the other, it is not reasonable to draw such a conclusion here either.

Another necessary point is that we do not want to say that under any circumstances, even in the presence of inconsistency, we must defend quantum mechanics. Firstly, it must be emphasized that the existence of two inconsistent propositions in a theory does not mean its inconsistency. Secondly, we do not believe that quantum mechanics is the ultimate theory describing the microscopic world. We do not want to say that the final word belongs to quantum mechanics, and we should not consider alternative theories. With various advancements in physics, mathematics, philosophy, and logic, we may witness quantum mechanics giving way to a more complete, precise, or revolutionary theory. Just as classical mechanics once gave way to quantum mechanics, we may witness another revolution in science and abandon quantum mechanics. Our main point is that we should not discard quantum mechanics due to inconsistency. If we establish the logical foundations of quantum mechanics on paraconsistent logic, we should no longer be concerned about inconsistency. This means that even in the presence of inconsistency in quantum mechanics, reasoning and analysis will still proceed meaningfully and logically. Just as this theory has been successful in empirical observations, it has also found a proper logical path. Such an approach enables precise and deep analysis in situations where the theory is inconsistent with classical intuition and logic, making it easier to understand complex phenomena.

\section{Objections in quantum mechanics and paraconsistent logic}

The formal structures of quantum mechanics and paraconsistent logic exhibit fundamental differences:
\begin{enumerate}
    \item \textit{Formal Structure}: Quantum mechanics and paraconsistent logic are underpinned by fundamentally different formal structures. Quantum mechanics is typically articulated using the mathematical language of linear algebra and complex numbers. In contrast, paraconsistent logic is a variant of symbolic logic governed by a unique set of rules. This divergence in formal structure necessitates substantial reinterpretation when one attempts to map one onto the other, potentially engendering inconsistencies.
    \item \textit{Predictive Power}: The predictive power inherent in quantum mechanics and paraconsistent logic exhibits substantial disparity. Quantum mechanics has demonstrated remarkable success in accurately predicting physical phenomena. Conversely, paraconsistent logic, serving as a tool for reasoning, does not proffer empirical predictions. This discrepancy in predictive power could be construed as a conflict between the two.
    \item \textit{Empirical Validation}: Quantum mechanics and paraconsistent logic have markedly different empirical groundings. Quantum mechanics is firmly anchored in empirical evidence and experimental validation. Paraconsistent logic, however, possesses a different kind of empirical grounding. While it can facilitate reasoning about empirical phenomena, it lacks empirical content in itself. This difference in empirical validation further accentuates the potential conflicts between paraconsistent logic and quantum entanglement.
\end{enumerate}

\section{Conclusion}
This study demonstrates how paraconsistent logic can effectively address the inherent inconsistencies within quantum mechanics, mainly through the lens of quantum entanglement. By incorporating paraconsistent logic, the paper provides a structured approach to understanding phenomena such as superposition and non-locality, which in the context of quantum mechanics refers to the instantaneous correlation between particles separated by large distances. These phenomena traditionally challenge the constraints of classical logic. The detailed exploration of entanglement through this non-classical logical framework clarifies the theoretical descriptions of these quantum phenomena and enhances our understanding of their foundational implications. This approach underscores the utility of paraconsistent logic in providing a coherent interpretation where classical logic may lead to paradoxes or inconsistencies, such as those illustrated by the EPR paradox and violations of Bell’s inequalities.

The conclusions drawn from this research not only invite further scrutiny and development within the field but also call for a collective effort to explore the potential of paraconsistent logic in quantum theory. While the study effectively bridges some gaps between quantum mechanics and logical consistency, it also opens up avenues for further research on the practical implications of these findings, particularly in how we might harness such insights in quantum computing or other technologies. However, it remains clear that while paraconsistent logic offers a novel perspective, adapting this logic into broader quantum mechanics practice and its potential to fundamentally alter quantum theory or technology still requires more profound exploration and empirical validation. This work lays a foundational step towards a more nuanced logical approach to quantum mechanics, encouraging ongoing dialogue and investigation into how inconsistencies are handled within the quantum domain, and inviting your valuable contributions to this exciting journey.

\section{acknowledgement}
We acknowledge the use of \href{https://quantum.ibm.com/}{IBM Quantum services} for this work. H. T. acknowledges the Quantum Communications Hub of the UK Engineering and Physical Sciences Research Council (EPSRC) (Grant Nos. EP/M013472/1 and EP/T001011/1).

\bibliographystyle{ieeetr}
\bibliography{sources}

\end{document}